\documentclass[prb,aps,twocolumn,showpacs]{revtex4}
\usepackage{graphics}
\usepackage{graphicx}
\begin{document}

\title{Energetics of hydrogen impurities in aluminum and their effect on
mechanical properties}

\author{
Gang Lu,
Daniel Orlikowski, Ickjin Park, Olivier
Politano and Efthimios Kaxiras} 
\affiliation{
Department of Physics and Division of Engineering and
Applied Sciences \\
Harvard University, Cambridge, MA 02138} 
\begin{abstract} 
\vskip 0.5cm
The effects of hydrogen impurities in the bulk and on the
surface of aluminum are theoretically investigated.  Within the
framework of density functional theory, we have obtained the
dependence on H concentration of the stacking fault energy,  
the cleavage energy, 
the Al/H surface energy and the Al/H/Al interface formation energy. 
The results indicate a strong dependence of the slip energy barrier
 in the $[\bar
211]$~direction the cleavage energy in the [111]~direction
and the Al/H/Al interface formation energy, on  
H concentration and on tension. The dependence of the Al/H surface energy on 
H coverage is less pronounced, while the optimal H coverage is $\leq 0.25$ 
monolayer.  
The calculated activation 
energy for diffusion between high symmetry sites in the bulk 
and on the surface is practically the same, 0.167 eV. 
From these results, we draw conclusions about the possible effect
of H impurities on mechanical properties, and in particular on
their role in embrittlement of Al.
\end{abstract}

\maketitle

\section{Introduction} 
In many technological applications of advanced materials a crucial
aspect of performance is the control of environmental effects, such as
the presence of impurities.  One such impurity is hydrogen, which
pervades most metals and degrades their performance.\cite{myers}
The interactions of H with lattice  
imperfections, such as dislocations, stacking faults, 
surfaces and microcracks, dominate its influence
on the mechanical properties of a material. However, these interactions 
are far less well understood at a fundamental level than the behavior of H in 
perfect crystals. Therefore, atomistic studies based on parameter-free,
{\it ab initio} calculations are of great interest because they can provide
accurate energetics for the various H-defect complexes and probe 
the microscopic physics responsible for the macroscopic 
behavior.   
The impurity-defect energetics are not only interesting by themselves, they can also be
incorporated in more sophisticated models in order to
make quantitative predictions for the macroscopic 
properties of solids, in what has become known as multiscale simulations of 
materials.\cite{gang2,ortiz}
 
The present study is motivated by the desire to shed light into  
H embrittlement of Al from an electronic structure point of view.  
Experimentally, the presence of H in
Al is associated with enhanced dislocation activity which, perhaps
paradoxically, leads to a brittle rupture failure.
\cite{myers,robertson,birnbaum} Theoretically, it has been
shown recently within the framework of the Peierls-Nabarro model,
that the presence of H in Al can dramatically enhance dislocation mobility 
and inhibit dislocation cross-slip.\cite{gang2} However, the underlying 
atomic bonding features that give 
rise to such dislocation behavior have not been explored. In this paper,
we show how H can change the nature of chemical bonding in Al,
leading to so-called hydrogen enhanced local plasticity (HELP). 
\cite{myers,gang2} Moreover, we show that the cleavage energy, 
which represents the ultimate resistance 
to fracture, can be considerably reduced by H. 

Another important aspect of H behavior in Al is  
the thermodynamics of H in bulk Al and on its (111) surface, and the corresponding
diffusion constants. The stability and mobility of H impurities in Al play an
important role in HELP and H embrittlement of Al. For example, it is 
observed experimentally that HELP occurs only when the thermal diffusion 
of H in the lattice is 
fast enough to follow the motion of dislocations.\cite{myers} Accordingly,
we calculate the diffusion energy barriers and the diffusion constants for H
in the bulk and on the (111) surface of Al. We also examine the H
diffusion process under uniform tensile strain to simulate the behavior at
the dislocation core or near the crack tip region.    

The outline of this paper is as follows: we
briefly describe our computational methodology in Sec.~II. In Sec.~III we present
the detailed results for the energetics of various defect structures involving
H in the bulk and on the (111) surface of Al and attempt to understand some of
the energetics from an electronic structure point of view. We discuss the
physical consequences of our results on the mechanical properties of Al and conclude
in Sec.~IV.

\section{Methodology}

The {\it ab initio} calculations we performed are based on
density functional theory with the VASP implementation \cite{vasp}
and ultra-soft pseudopotentials.\cite{vanderbilt} 
We have performed
$k$-point convergence studies in all cases using a uniform
Monkhorst-Pack scheme.\cite{monkhorst}  From these studies we have
determined that a grid consisting of 16$\times$16$\times$16 divisions
in the Brillouin-Zone of the primitive unit cell of bulk fcc Al,
appropriately scaled for larger unit cells, is adequate for good
convergence.  The kinetic energy cutoff of 130 eV for pure Al yields 
well converged results, whereas a higher cutoff of 350 eV is needed
in the presence of H atoms.
We have also introduced a smearing of the
Fermi surface by a temperature of 25~meV.  
With these computational parameters, the calculated lattice
constant for bulk fcc Al is $a = 3.99$ \AA~ and the bulk modulus is
$B = 83.2$ GPa, determined by a Birch-Murnaghan fit to the energy versus
volume curve.\cite{birch}  These values compare well with experimental
values \cite{hirth} of $a =4.05$ \AA\/ and $B = 76.9$ GPa, respectively.

\begin{figure}[!tp]
\includegraphics[width=230pt]{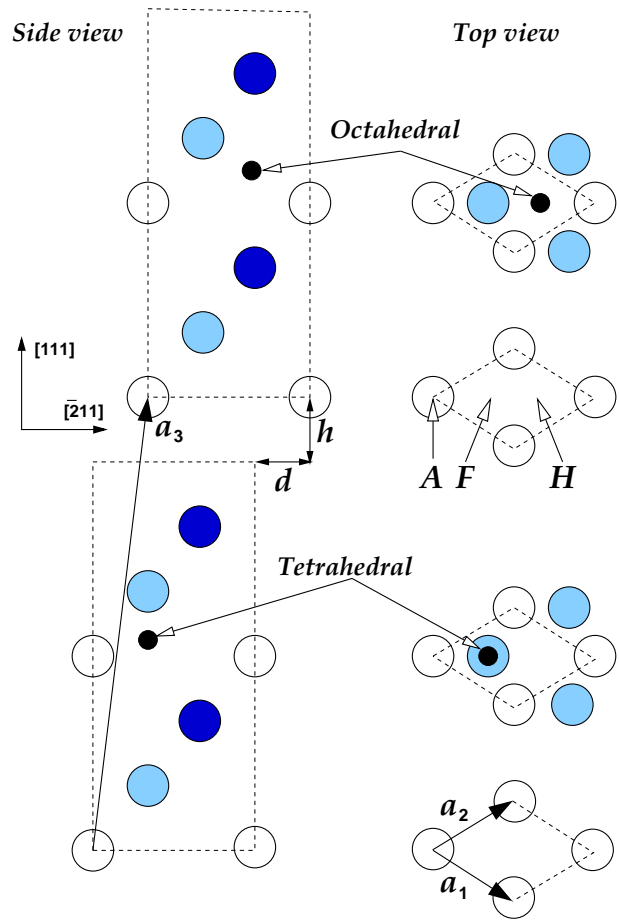}
\caption{Schematic representation of the geometry used in the total-energy
calculations. The basic supercell is shown in side and top views. 
The dashed lines outline the undistorted supercell. 
It consists of 6~layers in the [111]~direction and
has repeat vectors in the (111) plane equal to the ideal crystal
primitive lattice vectors, denoted by $\vec{a}_1, \vec{a}_2$. 
The third vector, $\vec{a}_3$, is along the 
[111] direction in the undistorted cell. 
Distortions of the cell by adding to $\vec{a}_3$ components 
in the $[{\bar 2}11]$ direction (denoted by
$d$, given in units of $a/\sqrt{6}$), or in the 
$[111]$ direction (denoted by $h$, given in units of $a/\sqrt{3}$),
lead to configurations relevant to the generalized stacking fault 
energy surface or to the cleavage energy; such a distortion 
for $d=1, h=1$ is illustrated.
The large white, grey and black circles indicate the positions of 
the Al atoms; all Al atoms contained in a unit cell are 
shown in the side view, but only selected planes of atoms are 
shown in the top view.  
The smaller black circle indicates the high 
symmetry positions of the H atom in the bulk (tetrahedral and 
octahedral), and on the 
(111) surface ($F$ for the fcc site, $H$ for the hcp site and $A$ for 
the atop site).
}
\label{setup}
\end{figure}

We next turn our attention to the atomic structures used to represent the
physical systems of interest.  In Fig.~\ref{setup} we show the basic 
supercell which consists of 6 Al layers in the [111] direction. 
In all the calculations, we used the same number of layers in this 
direction to represent the (111) surface or the interface
between two semi-infinite slabs.
We have used this supercell and multiples or distortions of it 
to calculate the generalized stacking fault energy curve and 
the surface and interface energies.
The unit cell with periodicity in the (111) plane equal to that of 
the bulk crystal will be referred to as the $1\times 1$ cell. 
Multiples of the in-plane vectors, denoted as $\vec{a}_1$ and $\vec{a}_2$ 
in Fig.\ref{setup},    
were used to create larger supercells 
for studying the effects of H concentration.
We have used $(q \vec{a}_1 \times q \vec{a}_2)$ multiples of the basic cell
with $q=1$, $\sqrt{3}$ and $2$. 
In each supercell we included one H atom. These configurations 
correspond to H concentrations in the bulk of 14.3,  
5.3 and 4.0 at.\%, respectively.  On the surface, they correspond to H  
monolayer (ML) coverages of $\Theta$ = 1.0, 0.333 and 0.25 ML, respectively.
We also report a single surface calculation with a $4\times 4$ unit 
cell, corresponding to H coverage $\Theta = 0.0625$ ML, 
in order to establish the value of the H/Al surface energy 
in the low coverage limit. 
The H atom in each supercell was placed at the high symmetry interstitial
sites, identified as the tetrahedral, 4-fold coordinated ($T$) or
octahedral, 6-fold coordinated ($O$) position in the undistorted bulk
configuration, or as the fcc ($F$), hcp ($H$)
and atop ($A$) sites on the (111) surface, all shown in Fig. \ref{setup}.  
Calculations with the H atom in positions between the high symmetry 
sites in the bulk and on the surface 
were used to obtain diffusion energy barriers.

Distorting the $q\times q$ supercell by increasing the horizontal
or vertical components of $\vec{a}_3$, the lattice vector which in 
the undistorted case lies along the [111] direction, 
produces configurations that generate the 
generalized stacking fault energy or the cleavage energy.
These two distortions are referred to as $d$ and $h$ respectively, 
and are given in their natural units of $a/\sqrt{6}$ and $a/\sqrt{3}$. 
In these units, $d = 1$ corresponds to the intrinsic stacking 
fault configuration, $d = 2$ corresponds to the so-called run-on 
configuration (in which two Al atoms on either side of the slip 
plane are exactly above and below each other), and $d = 3$ 
corresponds to another ideal configuration identical to $d = 0$.
Similarly, $h = 1$ 
corresponds to a separation between the two slabs 
equivalent to a missing
(111) layer.  

For each of these configurations, all atoms except the innermost
two layers of the Al slab
were fully relaxed via the conjugate-gradient method, so that the
magnitude of the calculated forces on the atoms was less than
0.03~eV/\AA.  
For the calculations of the energy barriers for diffusion, 
the coordinates of the H atom are held fixed, either in all three 
directions for bulk diffusion, or in the lateral surface directions   
for surface diffusion.
For the calculation of the generalized stacking fault energies
using the distorted bulk supercells (see below), 
the H atom was placed initially 
close to the interpolated tetrahedral or octahedral positions and allowed
to relax to the nearest local energy minimum.
We report energy differences between various configurations in eV
and surface energies in Jm$^{-2}$, in order to comply with 
conventions in the literature and make our results easily 
comparable to other published work.

\begin{figure}[tp]
\includegraphics[width=220pt]{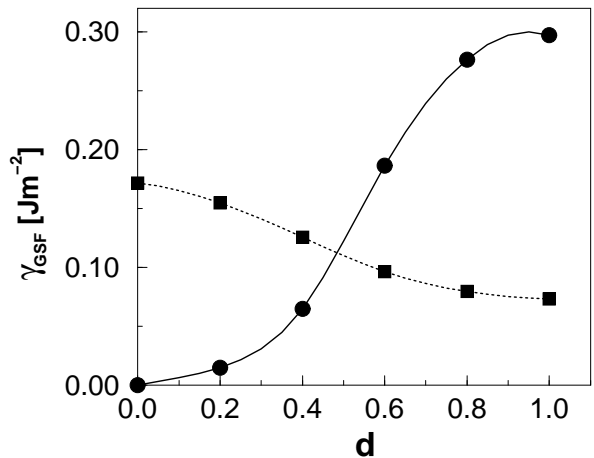}
\caption{
The generalized stacking fault energy $\gamma_{GSF}$
for Al with H~impurities at zero tension opening, $h=0$, as a
function of the slip $d$ in the $[\bar211]$~direction for the 
$1\times 1$ supercell, corresponding to H concentration of 14.3~at.\%:
circles represent the energy for the tetrahedral site, 
squares for the octahedral site 
(lines are fits intended as guide to the eye).  
}
\label{crossover}
\end{figure}

\section{Energetics of H~impurities in Al}

\subsection{Generalized stacking fault energies}
The generalized stacking fault~(GSF) energy, denoted by $\gamma_{GSF}$, 
is defined as the energy cost per unit area for sliding two
semi-infinite slabs relative to each other along a particular plane by
a certain vector $\vec{d}$. The energy surface generated by 
spanning the allowed values of $\vec{d}$ 
contains several important features relevant 
to the mechanical properties of solids and in particular to their 
brittle versus ductile behavior. 
For an fcc~metal like Al,
the most interesting portion of the GSF energy surface is the path
along the $[\bar211]$~direction on the (111)~plane.  
This path includes  both the
intrinsic stacking fault as well as the unstable stacking
fault, 
corresponding to $d = 1$ and $d = 0.6$, respectively.
The intrinsic stacking fault energy, denoted as $\gamma_{sf}$,
along with the elastic properties of the material, determine 
the separation distance between partial 
dislocations \cite{hirth} 
which controls the mobility of the dislocations.\cite{duesbery2}
The unstable stacking fault energy, denoted as $\gamma_{us}$, 
represents
the energy barrier for dislocation nucleation 
from a crack tip, which is related to the tendency for brittle
or ductile behavior of the material.\cite{rice}

The values of $\gamma_{GSF}$ for pure Al have been published 
elsewhere.\cite{gang,sun}  In the present work we have repeated these 
calculations to obtain a consistent set of numbers with the 
computational parameters and methodology adopted here.  The values
of the important configurations obtained by the present calculations are:
$\gamma_{us}=0.182$ Jm$^{-2}$, 
and $\gamma_{sf}$~=~0.134~Jm$^{-2}$, both for $h = 0$.
Including tension opening ($h \neq 0$), reduces these values 
dramatically. For example, for $h = 0.1$, we obtained
$\gamma_{us}=0.094$ Jm$^{-2}$, a 50\% reduction, 
and $\gamma_{sf}=0.092$ Jm$^{-2}$, a 30\% reduction. 

The values of 
$\gamma_{GSF}$ in the presence of H depend on the position 
of the H atom in the lattice and the H concentration. 
We next examine these two contributions separately.
We consider first our findings for the highest H concentration,   
14.3~at.\%, which corresponds to one H~atom in a
1$\times$1 supercell.  
The results, shown in
Fig.~\ref{crossover}, indicate that there is a cross-over at
approximately $d=0.5$ in site preference for the H~atom.  This
cross-over in site preference significantly reduces the 
unstable stacking energies to $\gamma_{us} = 0.097$ Jm$^{-2}$, a
$\sim$ 50\% reduction,
and the intrinsic stacking fault energy to $\gamma_{sf} = 0.073$
Jm$^{-2}$, also a $\sim$ 50\% reduction. 
The reason for such energy reduction is that the volume 
available for the interstitial H atom situated 
at the original
octahedral site decreases during the slip, while it increases 
at the tetrahedral site.
The effects of tension on $\gamma_{GSF}$ in the
presence of H were also calculated 
and found to be very similar to those for pure Al, as far as 
the relative energy decrease is concerned. 

\begin{table}[bp]
\caption{The unstable stacking energy, $\gamma_{us}$,
stacking fault energy, $\gamma_{sf}$, and cleavage energy, 
$\gamma_{cl}$, for the H/Al system as
a function of H~concentration without volume relaxation.
The ratio
$\gamma_{cl}/\gamma_{us}$ is also included.
}
\begin{ruledtabular}
\begin{tabular*}{\columnwidth}{@{\extracolsep{\fill}}c|c|cccc}   
Supercell & at.\% H & ${\gamma}_{us}$ 
& ${\gamma}_{sf}$ 
& $\gamma_{cl}$ & $\gamma_{cl}/\gamma_{us}$ 
\\
 & & (Jm$^{-2}$) & (Jm$^{-2}$) & (Jm$^{-2}$) & \\
\hline 
1$\times$1             & 14.3  & 0.097 & 0.073 &  0.930 & 9.6 
\\ 
$\sqrt{3}\times\sqrt{3}$&  5.3  & 0.089 & 0.071 & 1.611 & 18.1 
\\
2$\times$2             &  4.0  & 0.136 & 0.074 &  1.680 & 15.6 
\\      
\hline 
1$\times$1             &  0.0  & 0.182 & 0.134 & 1.934 & 10.6 \\
\end{tabular*}
\end{ruledtabular}
\end{table} 

To study the dependence of $\gamma_{us}$
and $\gamma_{sf}$ on H~concentration, we have computed these
energies for several H concentrations.
The results are given in Table~I. 
The general trend for both energies is to increase with 
decreasing H~concentration.  An exception to the trend is the 
highest H concentration at 14.3~at.\% (from the $1\times 1$ supercell),
which has {\em higher} values for $\gamma_{us}$ and $\gamma_{sf}$
than the next lower H concentration of 
5.3~at.\% (from the $\sqrt{3}\times\sqrt{3}$ supercell).
We believe that this has to do with the fact that in the $1\times 1$
supercell at fixed volume, the optimal ionic bonding 
distances
between H and Al atoms cannot be satisfied, and therefore
the system cannot attain a structure with a reasonably low energy.
To investigate this conjecture, 
we have also examined thoroughly the effect of volume
relaxation in the $1\times 1$ supercell. We find that 
in this supercell there is actually a low energy
configuration at the run-on position, $d=2$, which is
lower in energy by 0.615 Jm$^{-2}$ than the undistorted configuration 
of $d=0$, and involves an increase in the volume by 18\%. 
 In this configuration, the H atom
lies exactly between the two Al atoms on either side of the slip plane, 
forming strong ionic bonds across the interface, which 
compensates the energy loss due to the distortion of the Al lattice. 

\begin{figure}
\includegraphics[width=230pt]{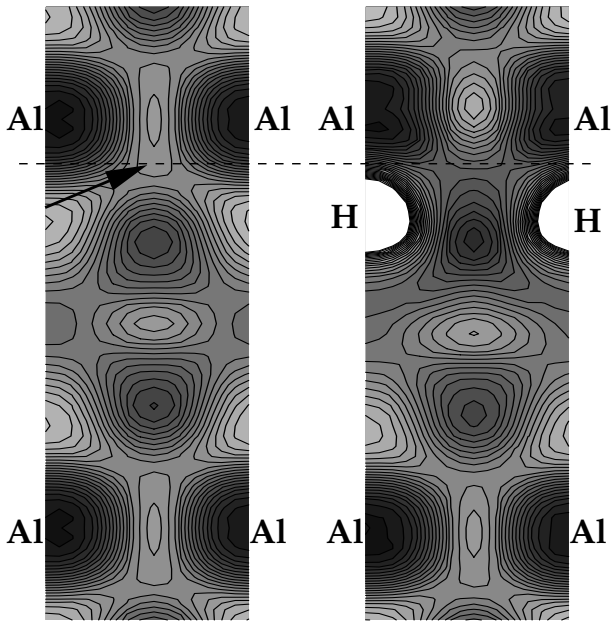}
\caption{The bonding valence charge density on the($\bar{1}2\bar{1}$)
plane for pure Al (left) and Al+H (right) systems at $d=0$. The horizontal
direction is [10$\bar{1}$] and the vertical direction is [111]. 
The fictitious slip plane is shown as a dashed line and the positions of
atoms are indicated by the corresponding labels. The arrow indicates
the buildup of valence charge density corresponding to covalent
bonding across the slip plane.}
\label{Contour}
\end{figure}

\subsection{Cleavage energy}
Another important energy for the mechanical behavior of a solid is the
cleavage energy, $\gamma_{cl}$, defined as the energy cost 
per unit area to separate the solid into two semi-infinite halves by  
creating two surfaces.  For the pure Al case, $\gamma_{cl} = 2\gamma_{s}$,
with $\gamma_{s}$ the energy of the newly created surface.  
We have calculated the cleavage energy with the various supercells
as the vertical component of $\vec{a}_3$ is increased up to $h=4$,  
corresponding to a separation of about 9 \AA. 
The H~atom was placed at the energetically
preferred tetrahedral site.  With the introduction of H to the
system, the cleavage energy dramatically decreases by as much as 50\%
at H concentration of 14.3~at.\% (see Table I). The decrease of the 
cleavage energy is approximately proportional to the H concentration.  

In order to elucidate the origin of the reduction in the GSF energy 
and the cleavage
energy in the presence of H, we examine the bonding
charge density on the ($\bar{1}2\bar{1}$) plane for pure Al and 
for Al+H (14.3 at.\%) at $d=0$, which is shown in Fig.~\ref{Contour}. 
The bonding charge
density is defined as the difference between the valence charge density
in the solid and the superposition of neutral atomic valence charge
densities placed at the lattice sites. The positive (negative)
bonding charge density represents the net gain (loss) of
charge as the atoms are brought together to form the solid. The
contour graph is shaded in such a way that regions with higher
value of charge density are lighter.
By examining the bonding charge density of Al  
and Al+H, we find extended covalent bonding (indicated by the arrow)
in Al across the slip plane, which is
dramatically weakened in the presence of H.
In fact, the H atom depletes the Al bonding charge from 
the interstitial region and the regions across the slip plane to form 
ionic bonding between the H sites and the nearest Al sites above it. 
As a consequence, the cohesive strength across the slip plane is reduced
by the presence of H, giving rise to the lower cleavage energies.
More importantly, since the strength of the ionic bonding
 between the positively
charged Al plane and negatively charged H plane is not sensitive to the 
relative sliding between the two planes, the sliding energy barrier 
is greatly 
reduced, and the GSF energy surface becomes much smoother 
\cite{gang2} in the presence of H. This is contrasted to the 
pure Al case, where the 
covalent bonding among Al atoms across the slip plane is very
sensitive to the local bonding distortions and consequently 
the GSF energy is higher and has more pronounced features. 
\cite{gang} Although these 
calculations concern Al,
we believe that the results are also applicable to other metals
whose electronegativity is lower than H. 

In Table~I we also give the
ratio of $\gamma_{cl}$ to $\gamma_{us}$ for the various
H~concentrations with pure Al as the reference point. The value of this
ratio is indicative of the tendency of the material to exhibit brittle
or ductile behavior.\cite{rice}  From this comparison we infer that at
modest H~concentrations the system has increased ductility,
which is consistent with the experimental observations 
concerning HELP,\cite{myers} but at the 
highest H~concentration considered the system may become less ductile.
The anomaly of the $\gamma_{cl}/\gamma_{us}$ ratio at the highest 
H concentration is related to the anomalous behavior of $\gamma_{us}$
at this concentration, as noted earlier. 

\subsection{Absorption and diffusion of H in bulk Al}
In view of the importance of the thermodynamics of H in bulk Al
and the kinetics of H transport in the presence of defects, 
we have performed additional calculations for the energetics of H 
absorption and the H diffusion energy barrier in
bulk Al. 
For these calculations we employed a 
32-atom supercell of the bulk crystal, which is a multiple of the 
conventional simple cubic cell by a factor of 2 in each direction
(hence, we refer to it as the 2$\times$2$\times$2 supercell).
We have investigated the two high symmetry
interstitial sites of a 
single H~impurity,  which corresponds to a H concentration of
3.03~at.\%.  The absorption energy, $E_{ab}$, was obtained with
reference to the cohesive energy 
of crystalline fcc~Al, $E_{c}$(Al), and using a gas of H$_2$ molecules, 
whose binding energy is $E_{b}$(H$_2$),  
as a reservoir for the H atoms:
\begin{equation}
E_{ab} = E_{c}({\rm H/Al}) - N_{sc} E_{c}({\rm Al}) -
\frac{1}{2} E_{b}( {\rm H}_2),
\label{E_ab}
\end{equation}
where $E_{c}$(H/Al) 
is the calculated cohesive energy of the Al supercell configuration
with one H impurity and $N_{sc}$ is the number of 
Al atoms in the supercell ($N_{sc}=32$ in the present case).  The
binding energy of the H$_2$ molecule was calculated to be $-6.697$ eV in
vacuum, using a cubic cell with side equal to 24 \AA, and the same 
computational parameters as for the H/Al system.
We find $E^{(T)}_{ab} = -0.222$ eV for the tetrahedral 
site and $E^{(O)}_{ab} = -0.152$ eV for the octahedral site. 
These results show that the incorporation of H in bulk Al,         
starting with an Al crystal and H$_2$ gas, is a
thermodynamically exothermic process. Therefore the H impurity is
thermodynamically
stable in bulk Al.

\begin{figure}
\includegraphics[width=240pt]{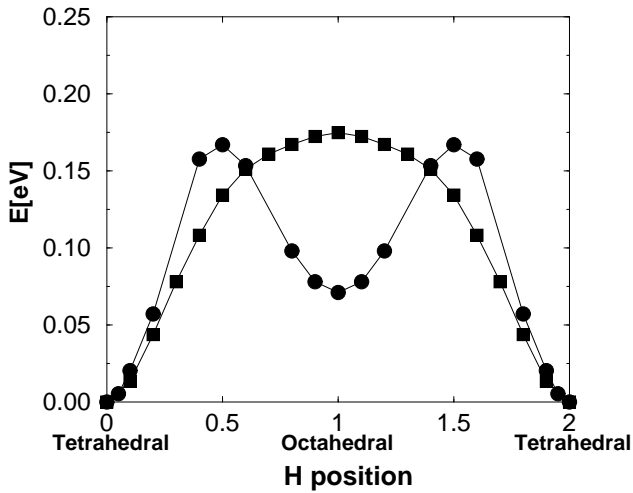}
\caption{The relative energy for motion of a H~atom from the
tetrahedral position (0) to the octahedral position (1) in the 
32-atom bulk supercell. Circles represent the energy without tensile strain,
and squares correspond the energy with 5\% tensile strain.}
\label{Act_bar}
\end{figure}

Having established the stability of H in bulk Al, we investigated
the mobility of H in the Al lattice. As alluded to earlier, the diffusion
rate of H in the lattice determines whether H-enhanced local plasticity 
can occur or 
not. HELP takes place only when the diffusion of H atoms 
is fast enough to allow them to redistribute
around the core of a moving dislocation (dynamic trapping) 
and thereby continuously minimize
the system energy.\cite{myers} H diffusion can also affect 
the kinetics of crack
propagation, the strain rate dependence of H embrittlement, 
and the rate of hydride formation.\cite{book1}
Since H jumps between nearby interstitial sites in bulk Al,
we calculated the activation energy of H diffusion 
between the closest tetrahedral and octahedral sites.  
In Fig.~\ref{Act_bar} we show the energy
as a function of relative position of the H~atom moving between
the tetrahedral~(0 and 2) and the octahedral site~(1). Another special
site is the mid-point between the tetrahedral and the octahedral sites.
We define the energy of the mid-point and of the octahedral
site relative to the tetrahedral site as $\varepsilon_{1}$ and 
$\varepsilon_{2}$, respectively.
The diffusion energy barrier for unstrained bulk Al
is $\varepsilon_{b}=0.167$ eV, 
which compares well with the experimental
value \cite{young} of 0.168~eV. The energy barrier configuration
happens to coincide with the mid-point between the octahedral
and tetrahedral sites in the unstrained crystal, that is 
$\varepsilon_{b} = \varepsilon_{1}$. 
Since we are also interested in the interaction of H with lattice
imperfections, such as dislocations, microcracks, etc., we
considered how H diffusion is affected by the presence of such defects.
A simple way to simulate this effect 
is to apply strain to the system. In this
study we concentrate on how tensile strain affects
H diffusion in Al, since this is the type of strain field
usually found around an edge
dislocation or a crack tip under model I loading, which are both relevant
to H embrittlement of Al. For the range of tensile strains studied,
the tetrahedral site has the lowest energy. When the hydrostatic 
tensile strain is
small ($\leq 3\%$), the energy barrier is located at the mid-point
between the tetrahedral and the octahedral sites, i.e., 
 $\varepsilon_{b} = \varepsilon_{1}$. 
Moreover, we find that for small strain
the diffusion energy
barrier drops monotonically as the strain increases, and is reduced to
0.150 eV for 3\% expansion. This result is 
important because it shows that not only H prefers to stay in
slightly enlarged interstitial regions such as dislocation cores and
crack tips, but that it can also move more easily within such regions. 
On the other hand, if the tensile strain is large ($> 3\%$),
the octahedral site becomes energetically unstable  
and represents the energy maximum where the energy barrier for diffusion is 
located, that is  $\varepsilon_{b} = \varepsilon_{2}$.
 One example for such diffusion energy profile 
is shown in Fig.~\ref{Act_bar}, 
corresponding to 5\% strain. The instability of the octahedral
site arises from the unfavorable bond length between H and Al 
atoms, 2.1 \AA, which is much larger than the preferred ionic bond
length of about 1.8 \AA.  
The values of $\varepsilon_{1}$ and $\varepsilon_{2}$ as a function
 of tensile strain are 
summarized in Table II. 
Noticing that the tetrahedral site
is always energetically favorable regardless of the strain,
we also calculated the relative energy of H at 
the mid-point of the direct line between two adjacent tetrahedral 
sites defined as $\varepsilon_{3}$ as a function of tensile 
strain. We find this energy difference
to be always
higher than the corresponding values of $\varepsilon_{1}$ 
and $\varepsilon_{2}$.
Therefore we have
confirmed that the 
tetrahedral-octahedral-tetrahedral sequence is the preferred diffusion
path for H in bulk Al.  
The various values of $\varepsilon_{3}$ 
as a function of tensile strain are also
listed in Table II.

The calculated energy profiles
allow us to estimate the bulk diffusion constant,
\begin{equation} 
D_{b} = \nu_{b} l_{b}^2 \exp[-\varepsilon_{b}/k_B T]
\end{equation}
where $\nu_b$ and $l_b$ are the attempt frequency and hopping length
for bulk diffusion. 
Approximating the energy differences near the equilibrium 
tetrahedral configuration 
by a second order polynomial in the distance, we find an attempt 
frequency $\nu_{b} = 0.8 \times 10^{11}$ sec$^{-1}$, while the hopping
length between equivalent sites is $l_{b} = 0.948~a$. The
diffusion constant at room temperature (300 K) is estimated
to be $1.78 \times 10^{-11}$ m$^2$s$^{-1}$ in the unstrained
crystal. Assuming that the values of the attempt frequency and
hopping length are not significantly affected by strain, the
value of the bulk diffusion constant at room temperature and
for 3\% and 5\% tensile strain is $3.46 \times 10^{-11}$ m$^2$s$^{-1}$
and $1.31 \times 10^{-11}$ m$^2$s$^{-1}$, respectively.

\subsection{Adsorption and diffusion of H on Al(111)}
For H embrittlement of metals, it has been experimentally observed 
that the fracture surface is 
along the slip plane, where shear localization occurs.\cite{myers,
matsumoto} Apparently, adsorption and diffusion of 
H on the fresh fracture 
surface play an important role on the kinetics of
crack propagation and the embrittling effect of H.
The critical energetics that are relevant to the adsorption and diffusion
process not only can provide insight into the problem, they can 
also be used in an empirical analysis which can deal with the macroscopic 
aspects of this phenomenon.\cite{ortiz} 

\begin{table}[bp]
\caption{The energy of special points for H diffusion in bulk Al,
relative to the tetrahedral position: $\varepsilon_{1}$ is the
energy of the mid-point between the tetrahedral and octahedral
positions, $\varepsilon_{2}$ is the energy of the octahedral position
and $\varepsilon_{3}$ is the energy of the mid-point between adjacent
tetrahedral sites. The asterisks denote the energy barrier for diffusion.} 
\begin{ruledtabular}
\begin{tabular*}{\columnwidth}{@{\extracolsep{\fill}}cllc} 
\multicolumn{1}{c}{Strain}&
\multicolumn{1}{c}{$\varepsilon_{1}$}&
\multicolumn{1}{c}{$\varepsilon_{2}$}&
\multicolumn{1}{c}{$\varepsilon_{3}$} \\
\multicolumn{1}{c}{(\%)}& 
\multicolumn{1}{c}{(eV)}& 
\multicolumn{1}{c}{(eV)}& 
\multicolumn{1}{c}{(eV)} \\
\hline
0  &  0.167 * &0.071  &0.359 \\
1  &  0.161 * &0.094  &0.332 \\
2  &  0.156 * &0.116  &0.306 \\
3  &  0.150 * &0.137  &0.280 \\
4  &  0.143 & 0.157 * &0.255 \\
5  &  0.134 & 0.175 * &0.229 \\
\end{tabular*}
\end{ruledtabular}
\end{table}

\begin{figure}
\includegraphics[width=230pt]{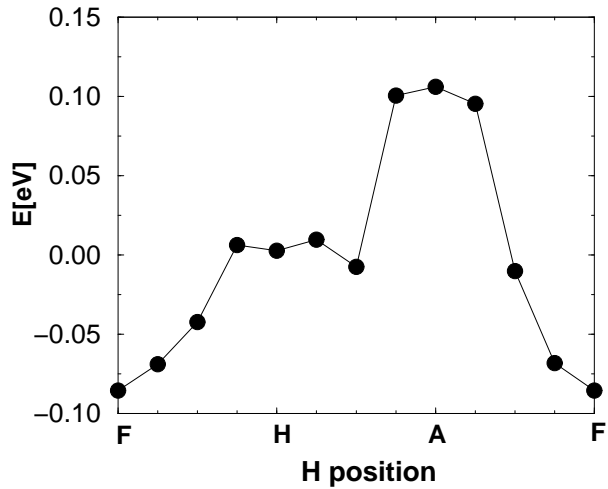}
\caption{The adsorption energy for H on the (111)~Al surface
within the 1$\times$1~surface unit cell corresponding to
H coverage $\Theta=1$ ML, for different positions in the $[{\bar 2}11]$
direction: $F=$ fcc site, $H=$ hcp site, $A=$ atop site.}
\label{Surf_ab}
\end{figure}

For this reason, we have considered the adsorption of a H atom
on the (111)~surface of Al, in configurations corresponding to 1$\times$1,
$\sqrt{3}\times\sqrt{3}$, and 2$\times$2 surface unit
cells, or H coverages in the range $\Theta \in$ [0.25, 1.0] ML. 
The H atom was placed at the high-symmetry $F, H$ and $A$ points
(see Fig. \ref{setup}), as well 
as points at regular intervals between them, to determine the lowest 
energy configuration and the energy barrier for surface diffusion.   
By analogy to the definition of Eq.(\ref{E_ab}), 
the adsorption energy $E_{ad}$ is defined as the energy of the 
configuration with a H atom on the (111) Al surface, relative to 
the same surface without H and using a gas of H$_2$ molecules
as a reservoir for the H atoms. 
For a full
monolayer of H ($\Theta$ = 1.0) on the (111)~Al surface, 
corresponding to the $1\times 1$ cell, the
adsorption energy for the various positions along the
$[\bar211]$~direction is given in Fig.~\ref{Surf_ab}.  The
fcc~site~($F$) is the energetically preferred position with 
$E_{ad}=-0.085$ eV, whereas the hcp~site~($H$) has nearly zero
adsorption energy, and the atop site~($A$) is energetically
unfavorable. We find that when displacing the H atom from the $F$ toward 
the $H$ and $A$ positions,  
the Al~atoms near the surface are also displaced in the
$[\bar 211]$ direction, so as to maintain the high 
coordination of the H~atom to the extent possible. 
This is especially pronounced at the bridge position, which is
half way between the $F$ and the $H$ sites (see Fig.~\ref{setup}). 

From the results of these calculations we conclude that the 
diffusion of H on the (111) Al surface will follow a zig-zag 
path between successive $F$ and $H$ sites. 
We find that the energy difference between the $H$ and $F$ sites 
is a reasonable
approximation for the diffusion energy barrier, $\varepsilon_{s}$, 
within the numerical uncertainty inherent in the calculations. 
With this in mind, we have calculated the diffusion energy barrier 
as a function of H coverage, 
using multiples of the $1\times 1$ surface unit cell
which is given in Table III.
It is clear from this Table
that the energy barrier for surface diffusion of an isolated 
H atom is approximately 0.163 eV, practically the same as that for bulk diffusion.  
We can also obtain an estimate of the surface diffusion constant 
using the energy as a function of H position from the calculation
of the $1\times 1$ unit cell: 
\begin{equation} 
D_{s} = \nu_{s} l_{s}^2 \exp[-\varepsilon_{s}/k_B T]
\end{equation}
where $\nu_s$ and $l_s$ are the attempt frequency and hopping length
for surface diffusion.  Using the same procedure as for bulk diffusion,  
we find an attempt 
frequency $\nu_{s} = 0.7 \times 10^{10}$ sec$^{-1}$, the hopping
length between equivalent sites is $l_{s} = 0.707~a$, and the
diffusion constant at room temperature (300 K) is $1.02 \times 10^{-12}$
m$^2$s$^{-1}$.                 

Finally, we note that the presence of H on the (111) Al surface 
reduces the surface energy considerably.  In order to quantify 
this observation, we report in Table III the calculated adsorption
energy, $E_{ad}$, for H atoms at various coverages.  
The reduction in surface 
energy in the presence of H, $\Delta\gamma_{s}$,  
is obtained by converting the adsorption energy 
to a surface energy and subtracting from it the 
corresponding surface energy of pure Al. We find that 
this reduction in surface energy is a function of 
coverage, and it increases as the coverage decreases
(Table III). Finally
we have performed one additional calculation
in a larger 4$\times$4 surface unit cell in order to determine to what 
extent this trend continues for lower coverages; the result is
included in Table III. 
We conclude that 
the H coverage which gives the largest reduction in surface energy 
is in the range $0.0625 \leq \Theta \leq 0.25$ ML. 

It should be
pointed out that $\Delta\gamma_{s}$ is a very important material
parameter in determining the tendency of impurity-induced 
intergranular fracture. More specifically, according to the thermodynamic
theory developed by Rice and Wang,\cite{wang} the potency of a segregating
impurity in reducing the Griffith work of a brittle grain boundary
separation is a linear function of the difference
$\Delta\gamma_{gb} - \Delta\gamma_{s}$, that is, the difference between
the segregation energy of that impurity at a grain boundary and at
a free surface. A smaller reduction in surface energy
(a less negative $\Delta\gamma_{s}$) indicates a weaker tendency
for brittle intergranular fracture. Based on the fact that our 
calculated $\Delta\gamma_{s}$ is more than an order magnitude smaller than 
typical values for intergranular fracture,\cite{wang,lu,wu} we  
infer H-induced fracture in Al to be of a transgranular
nature. Of course,
the definite determination of the tendency will also depend on  
$\Delta\gamma_{gb}$, which is not available to us.

It is also instructive to compare this reduction in surface energy
to the energy required to form Al/H/Al interface structures
corresponding to the bulk unit cells with one H impurity 
discussed earlier.  In Table III we give the absorption energy $E_{ab}$
obtained from the corresponding unit cells employed in the bulk calculations. 
As before, we use the pure Al bulk as a reference 
system and a gas of H$_2$ molecules
as the reservoir for H atoms.
The configurations used for these calculations 
correspond to the formation of 
a planar interface between two (111) planes of bulk Al separated by an ordered 
layer of H atoms.  
We have calculated the interface formation energy,  
$\Delta \gamma_{int}$, as a function of H content
at the interface, expressed in ML of H; the results are given in Table III.
Consistent with the calculations we presented earlier, this formation energy 
is positive for large H concentrations (recall the large positive 
absorption energy for a H atom in the 1$\times$1 bulk supercell).
However, with decreasing H concentration at the interface we expect that this 
formation energy will be reduced, and for small enough 
concentrations it should be negative,  
corresponding to  
the absorption energy for an isolated H atom in
bulk Al, which we found to be $-0.222$ eV.  Indeed, 
$\Delta\gamma_{int}$ becomes negative for the 
2$\times$2 supercell.  Note that in this supercell the 
shortest distance between H atoms on the (111) plane
is $a\sqrt{2}$,
which is shorter than the distance between the H impurities in the 
32-atom bulk supercell, equal to $2a$.  

\begin{table}[bp]
\caption{
The energy difference 
between the two high symmetry positions $F$ and $H$ 
of a H atom on the Al(111) surface, identified as the 
surface diffusion activation energy $\varepsilon_{s}$; 
the adsorption energy $E_{ad}^{(F)}$ of H atoms 
at the energetically preferred fcc ($F$) site 
on the (111) surface and the corresponding  
reduction in surface energy $\Delta\gamma_{s}$; 
the absorption energy $E_{ab}^{(T)}$ of H atoms at 
the energetically preferred tetrahedral ($T$) site
in the interface between two 
(111) planes and the corresponding interface formation energy  
$\Delta\gamma_{int}$.
All quantities are given as functions of H coverage $\Theta$ in monolayers (ML).
The last line gives the corresponding results for the 32-atom bulk supercell
of the conventional cubic cell.
}
\begin{ruledtabular}
\begin{tabular*}{\columnwidth}{@{\extracolsep{\fill}}clccccc}  
Supercell &
$\Theta $ 
& $\varepsilon_{s}$ 
& $E_{ad}^{(F)}$  
& $\Delta\gamma_{s}$ 
& $E_{ab}^{(T)}$  
& $\Delta\gamma_{int}$ 
\\
& (ML) & (eV) & (eV) & (Jm$^{-2}$) & (eV) & (Jm$^{-2}$) \\
\hline 
1$\times$1             & 
1.0 & 0.092 & $-0.085$ & $-0.138$ & $+0.383$ & $+0.586$ \\ 
$\sqrt{3}\times\sqrt{3}$&  
0.333 & 0.157 & $-0.365$ & $-0.187$ & $+0.067$ & $+0.034$ \\
2$\times$2             &
0.25 & 0.163 & $-0.489$ & $-0.188$ & $-0.039$ & $-0.015$ \\      
4$\times$4             &
0.0625 & & $-1.907$ & $-0.183$ & &   \\      
\hline
2$\times$2$\times$2 & & 0.167 & & & $-0.222$ &  \\ 
\end{tabular*}
\end{ruledtabular}
\end{table} 

\section{Discussion and Conclusions}

The following picture emerges from the calculations reported above.
In a system consisting of a crystalline Al phase in equilibrium 
with a H$_2$ gas, H atoms will be adsorbed on
the (111) Al surfaces, the natural cleavage planes, in order to 
lower the surface energy.  The equilibrium coverage will be in the 
range $0.0625 \leq \Theta \leq 0.25$ ML.  Diffusion of H on the surface 
and in the bulk is relatively fast. 
Using the values for the diffusion constants we determined earlier, we find    
that at room temperature the length scale for surface diffusion in a time
 interval of 1 sec is 
$\sim 0.3~\mu$m, while for bulk diffusion it is $\sim 7~\mu$m. 
Over such distances the equilibrium surface  
H coverage and bulk H content will be dictated 
by thermodynamic considerations.  
The calculated bulk absorption energy for 
the tetrahedral site, 
$E^{(T)}_{ab} = -0.222$ eV, 
indicates that 
energetically it is possible for H to end up in the bulk.
However, this is not the preferred configuration.  
In fact, the calculated 
surface adsorption energies per H atom are lower than the bulk 
absorption energies. For example, from the calculation of 
the $4\times 4$ surface
unit cell with one
H atom at the $F$ position, we find a surface adsorption energy 
$E^{(F)}_{ad} = -1.907$ eV, significantly lower than the bulk
absorption energy.  In this configuration the H atom 
can be considered as an isolated atom on the Al surface.
It appears from these calculations that if there is any H in bulk Al
and if the system is allowed to equilibrate with the surface, H 
tends to diffuse out and remains on the surface.  
This conclusion rests on 
the assumption that a clean and atomically flat 
(111) Al surface is available, which is usually not the case in reality 
due to the tendency of Al surfaces to oxidize.  The presence
of an oxide on the surface will completely alter the thermodynamic
balance, making it possible for H atoms to remain in the bulk. 
Moreover, we have not investigated the possibility of surface vacancies 
or other defects on the surface, which can also change the thermodynamic picture.  
The presence of such defects can only lower the Al surface energy, 
while the binding of H atoms at such defects may not be preferred over 
binding on the flat surface, as the values of $E_{ab}^{(T)}$ and $E_{ad}^{(F)}$
reported in Table III suggest.  Thus, the presence of defects on the 
Al surface may also suffice to make the incorporation of H atoms 
in the bulk thermodynamically stable.

On the other hand, 
our calculations for H in bulk Al show that H atoms feel an effective 
repulsive interaction for distances shorter than twice the 
primitive lattice constant, $a/\sqrt{2}$.  
This is evident from the Al/H/Al interface energies
$\Delta\gamma_{int}$, reported in Table III, which are all
positive except for the 2$\times$2 supercell.  Therefore, H atoms in bulk
Al cannot form dense clusters but have to be apart from each other by a 
distance at least $a/\sqrt{2}$. This result is significant because
it spells out the importance of
H-H interactions and casts doubt on studies that ignore such 
interactions. 
The energetically favorable 
configuration which we found for high H concentrations
in the bulk, 
with a H atom between two Al atoms directly above and below it in 
the [111] direction, also has interesting implications.  
Since this structure has a lower energy than the undistorted crystal
configuration with equal H concentration but also involves a large 
volume relaxation (18\%), we conclude that if there are voids 
or other defects which give rise to tensile strain, 
H atoms will be preferentially 
bound to those sites, such as the cores of edge dislocations.
In fact, a recent study 
showed that the binding 
energy of H to the core of an edge dislocation is much larger
than that of a screw dislocation, \cite{gang2} which is in line 
with our observation
here. These results have an important
consequence for dislocation
motion: the edge dislocation needs to turn into a screw dislocation
in order to cross-slip, a process that will be hindered by
the binding of H atoms to the edge dislocation, 
consistent with experimental observations.\cite{ferreira} 
More importantly, this H-inhibited cross-slip will give rise
to slip planarity and possibly, shear localization, the two
most important elements to understand H embrittlement in terms
of the HELP mechanism.\cite{myers}

Finally, we turn our attention to the effects of H on the intrinsic stacking 
fault energy $\gamma_{sf}$, unstable stacking fault energy  
$\gamma_{us}$, and cleavage energy, $\gamma_{cl}$.  As already noted,
the presence of H reduces all these quantities relative to their 
values in pure Al.  The ratio $\gamma_{cl}/\gamma_{us}$ as 
a function of impurity content has been 
employed to discern brittle versus ductile response.\cite{duesbery}  
The simple physical picture behind this argument is that a low 
value of this ratio indicates a preference for cleavage rather than 
dislocation generation at a crack tip which is controlled by the
value of $\gamma_{us}$;\cite{rice} this behavior is associated with 
 brittle failure. Conversely, a high value of this ratio indicates 
the preference for dislocation generation at a crack tip, 
a behavior associated with ductile response.   
While this picture may be overly simplistic for quantitative analysis,
it does give insight about general trends and can even lead 
to useful predictions that have been verified experimentally.\cite{duesbery}
For H in bulk Al, this ratio is higher than the value in pure Al 
except at the highest H concentration considered here
(14.3 at \%, corresponding to the 1$\times$1 supercell),
clearly suggesting that H embrittlement in Al takes place 
as plastic
rupture rather than as brittle separation.\cite{myers} 
In fact, recent work 
\cite{gang2} indicates that H in Al can indeed lead to enhanced local 
plasticity, a precursor to H embrittlement.\cite{myers}   

The lower value of the intrinsic stacking fault energy in the presence 
of H suggests a larger 
separation of the partial dislocations in Al, which could hinder 
the dislocation cross-slip since the partial dislocations must be
constricted  
before cross-slip can take place. 
But a more detailed analysis based on the
Peierls-Nabarro model \cite{gang2} shows that, 
even though the stacking fault energy is lowered by the 
presence of H, the partial dislocations are not split any 
further than in pure Al, while the core width of the dislocations
is increased significantly giving rise to enhanced dislocation
mobility.

To summarize, we have performed density functional theory calculations
to study the energetics of
H impurities in bulk 
and on the (111) surface of Al.
We have obtained the
dependence of the stacking fault energy and the cleavage energy, as
well as the Al/H surface energy and the Al/H/Al interface formation energy, 
on H concentration. The results indicate that there is
a strong dependence of the GSF energy in the $[\bar
211]$~direction, the cleavage energy in the [111]~direction
and the Al/H/Al interface formation energy, on  
H concentration and on tensile strain. We are able to explain
the H-induced reduction of the stacking fault energy and cleavage
 energy in Al from an
electronic structure point of view, and conjecture that such reduction
can also take place in other H-metal systems.
It is found that the dependence of the Al/H surface energy on the
H coverage is less pronounced, with the optimal coverage being $\leq 0.25$ 
monolayer.  
The calculated activation 
energy for diffusion between high symmetry sites in the bulk 
and on the surface is practically the same, 0.167 eV, in good
agreement with experimental measurements. 
Although our calculations reported here provide strong theoretical 
evidence for the HELP
mechanism, they are not able to answer 
how HELP eventually leads to H embrittlement. 
Nevertheless, we believe that our work
sets the stage for developing a comprehensive theory of H embrittlement, 
which most
likely will necessitate a multiscale framework. 
\section{Acknowledgement}
We would like to thank Nick Choly, Emily Carter, Rob 
Phillips, and Bill Curtin for useful discussions.
We acknowledge the use of computer facilities at Brown University. 
This work was funded by AFOSR, Contract \# F49620-99-1-0272.


\begin{thebibliography}{99}

\bibitem{myers} S.M. Myers {\it et al}., Rev. Mod. Phys. {\bf 64}, 559
(1992) and references therein.

\bibitem{gang2} G. Lu, Q. Zhang, N. Kioussis, and E. Kaxiras, Phys. 
Rev. Lett., {\bf 87}, 095501 (2001).

\bibitem{ortiz} O. Nguyen and M. Ortiz (to be published).

\bibitem{robertson} I.M. Robertson, Eng. Frac. Mech. {\bf 64}, 649
(1999).

\bibitem{birnbaum} H.K. Birnbaum and P. Sofronis, Mat. Sci. Eng. A
{\bf 176}, 191 (1994).

\bibitem{vasp} G. Kresse and J. Furthm\"uller, Phys. Rev. B {\bf 54},
11169 (1996).

\bibitem{vanderbilt} D. Vanderbilt, Phys. Rev. B {\bf 41}, 7892 (1990).

\bibitem{monkhorst} H.J. Monkhorst and J.D. Pack, Phys. Rev. B {\bf
13}, 5188 (1976).

\bibitem{birch} F. Birch, J. Geophys. Res., {\bf 83}, 1257 (1978).

\bibitem{hirth} J.P. Hirth and J. Lothe, {\it Theory of
Dislocations}, Wiley: New York (1992).

\bibitem{duesbery2} M.S. Duesbery, Modelling Simul. Mater. Sci. Eng.
{\bf 6}, 35 (1998).

\bibitem{rice} J.R. Rice, J. Mech. Phys. Solids {\bf 40}, 239 (1992);
J.R. Rice, G.E. Beltz, and Y. Sun, {\it Topics in Fracture and
Fatigue}, ed. A. Argon (Berlin: Springer 1992).

\bibitem{gang} G. Lu, N. Kioussis, V.V. Bulatov, and E. Kaxiras,
Phys. Rev. B, {\bf 62}, 3099 (2000).

\bibitem{sun} Y. Sun and E. Kaxiras, Philos. Mag. A {\bf 75}, 1117 (1997).

\bibitem{book1} M. Hashimoto and R.M. Latanision, in {\it Chemsitry
and Physics of Fracture}, edited by R.M. Latanision and R. Jones 
(Martinus Nijhoff Publishers, Dordrecht, The Netherlands, 1987), p. 505. 
\bibitem{young} G.A. Young and J.R. Scully, Acta Mater., {\bf 46},
6337 (1998).

\bibitem{matsumoto} T. Matsumoto and H.K. Birnbaum, Trans. Jpn. Inst.
Met. {\bf21}, 493 (1980); S.P. Lynch, J. Mater. Sci. {\bf21}, 692 (1986).

\bibitem{wang} J.R. Rice and J.S. Wang, Mater. Sci. Eng. A {\bf107}, 23 (1989).

\bibitem{lu} G. Lu, N. Kioussis, R. Wu and M. Ciftan, Phys. Rev. B {\bf 59},
891 (1999).

\bibitem{wu} R. Wu, A.J. Freeman, and G.B. Olson, Science {\bf 265}, 376 (1994).

\bibitem{ferreira} P.J. Ferreira, I.M. Robertson and H.K. Birnbaum,
Acta Mater. {\bf 47}, 2991 (1999).

\bibitem{duesbery} 
U.V. Waghmare, E. Kaxiras, V. Bulatov and M.S. Duesbery,
 Modelling Simul. Mater. Sci. Eng.
{\bf 6}, 483 (1998); 
U.V. Waghmare, V. Bulatov, E. Kaxiras, M.S. Duesbery,
Mat. Sci. and Engin. A {\bf 261}, 147 (1999);
U.V. Waghmare, E. Kaxiras, V. Bulatov and M.S. Duesbery,
Philos. Mag. A {\bf 79}, 655 (1999);
U.V. Waghmare, E. Kaxiras and M.S. Duesbery,
Phys. Stat. Sol. B {\bf 217}, 545 (2000).

\end{thebibliography}
\end{document}